\documentclass[prd,preprint]{revtex4}
\usepackage{graphicx}
\usepackage{latexsym}
\usepackage{amsmath}
\newcommand{\bq}{\begin{equation}}
\newcommand{\eq}{\end{equation}}
\newcommand{\bqn}{\begin{eqnarray}}
\newcommand{\eqn}{\end{eqnarray}}
\newcommand{\lb}{\label}
\begin{document}
\title{Thermodynamical properties of the Universe with dark energy}
\author{Yungui Gong}
\email{yungui_gong@baylor.edu}
\affiliation{ CASPER, Department of Physics,
Baylor University, Waco, TX 76798, USA }

\author{Bin Wang}
\email{wangb@fudan.edu.cn}
\affiliation{Department of Physics, Fudan University,
Shanghai 200433, China}

\author{Anzhong Wang}
\email{anzhong_wang@baylor.edu}
\affiliation{CASPER, Department of Physics, Baylor University,
Waco, TX 76798, USA}
\begin{abstract}
We have investigated the thermodynamical properties of the
Universe with dark energy. Adopting the usual assumption in
deriving the constant co-moving entropy density that the physical
volume and the temperature are independent, we observed some
strange thermodynamical behaviors. However, these strange
behaviors disappeared if we consider the realistic situation that
the physical volume and the temperature of the Universe are
related. Based on the well known correspondence between the
Friedmann equation and the first law of thermodynamics of the
apparent horizon, we argued that the apparent horizon is the
physical horizon in dealing with thermodynamics problems. We have
concentrated on the volume of the Universe within the apparent horizon
and considered that the Universe is in thermal equilibrium with
the Hawking temperature on the apparent horizon. For dark energy
with $w\ge -1$, the holographic principle and the generalized
second law are always respected.
\end{abstract}
\preprint{gr-qc/0610151}
\maketitle
\section{Introduction}

Since the great discovery that the Universe is experiencing
accelerated expansion driven by dark energy (DE) from the type Ia
supernova (SN Ia) observations in 1998 \cite{riess}, many works have
been done in pursuit of understanding this spectacular phenomena.
The later more accurate SN Ia data \cite{sn}, together with the
Wilkinson Microwave Anisotropy Probe data \cite{wmap} and the Sloan
Digital Sky Survey data \cite{sdss} indicate that the Universe is
almost spatially flat and further support the existence of DE which
contributes about 72\% of the matter content of the present universe
within the framework of Einstein's general relativity. Despite the
robust observational evidence for the existence of DE, DE is still a
major puzzle of modern cosmology. Except knowing that DE has
negative pressure, we know little about its theoretical nature and
origin. In this work we would like to study the DE from
thermodynamical considerations.

In the semiclassical quantum description of black hole physics, it
was found that black holes emit Hawking radiation with a temperature
proportional to their surface gravity at the event horizon and they
have an entropy which is one quarter of the area of the event
horizon in Planck unit \cite{hawking75}. The temperature, entropy
and mass of black holes satisfy the first law of thermodynamics
\cite{hawking73}. On the other hand, it was shown that the Einstein
equation can be derived from the first law of thermodynamics by
assuming the proportionality of entropy and the horizon area
\cite{ted}. The Einstein equation for the nonlinear gravitational
theory $f(R)$ was also derived from the first law of thermodynamics
with some non-equilibrium corrections \cite{eling}. For a general
static spherically symmetric space-time, Padmanabhan showed that
the Einstein equation at the horizon gives the first law of
thermodynamics on the horizon \cite{padmanabhan02}. The study on the
relation between the Einstein equation and the first law of
thermodynamics has been generalized to the cosmological context
where it was shown that the first law of thermodynamics on the
apparent horizon $\tilde{r}_A$ can be derived from the Friedmann
equation and vice versa if we take the Hawking temperature
$T_A=1/2\pi\tilde{r}_A$ and the entropy
 $S_A=\pi\tilde{r}_A^2/G$ on the apparent horizon \cite{cai05}.
Furthermore, the equivalence between the first law of thermodynamics and
Friedmann equation was also found for gravity with Gauss-Bonnet term
and the Lovelock gravity theory \cite{cai05,cai06}.
These results disclosed that there may be a deep relationship between
the thermodynamics of the horizon and the Einstein equation, which might
shed some light on the properties of DE.

Motivated by the black hole thermodynamics, Bekenstein
proposed a universal entropy bound for a weakly self-gravitating physical system
in 1981 \cite{bekenstein}.
't Hooft and Susskind subsequently built an influential holographic principle \cite{holography}
relating the maximum number of degrees of freedom in
a volume to its boundary surface area \cite{bousso,adscft}. The extension of the
holographic principle to a general cosmological setting
was first addressed by Fischler and Susskind
\cite{fischler} and subsequently got modified by many
authors \cite{more,bak}. The idea of holography is
viewed as a real conceptual change in our thinking about gravity \cite{8}.
There have been some examples of applying the holographic principle to
understand cosmological problems. It is interesting to note that
holography implies a possible value of the cosmological constant
in a large class of universes \cite{9}.
In an inhomogeneous cosmology holography was also realized as
a useful tool to select physically
acceptable models \cite{6}. The idea of holography has further
been applied to the study of inflation
and gives possible upper limits to the number of e-folds \cite{10}.
Recently, holography has again been
employed  to investigate the DE \cite{11} and it is expected that the holographic
principle could serve as an effective way to help us understand the DE.

The Universe can be considered as a thermodynamical system. The
thermodynamics in de Sitter space-time was first investigated by
Gibbons and Hawking in \cite{gibbons}. In a spatially flat de
Sitter space-time, the event horizon and the apparent horizon of
the Universe coincide and there is only one cosmological
horizon. It was found that the area of the cosmological horizon
can be interpreted as the entropy or lack of information of the
observer about the regions which he or she cannot see, and an observer
with a particle detector will observe a background of thermal
radiation coming from the cosmological horizon. The
thermodynamical study of the Universe has been extended to the
quasi-de Sitter space in \cite{pollock,frolov,wang06}. When the
apparent horizon and the event horizon of the Universe are
different, it was found that the first law and the second law of
thermodynamics hold on the apparent horizon, while break down if
we consider the event horizon \cite{wang06}. Due to this subtlety,
we will concentrate our attention on the thermodynamical
properties of the Universe enveloped by the apparent horizon. We
will derive the temperature and entropy of the matter contents inside
the apparent horizon from the first law of thermodynamics and
discuss the holographic entropy bound and the generalized second
law (GSL) of thermodynamics for the Universe with DE.

In the discussion of the thermodynamical properties of the matter contents,
it is usually assumed that the physical volume and temperature of
the Universe are independent and
by using the integrability condition
$\partial^2 S/\partial T\partial V=\partial^2 S/\partial V\partial T$
one has the constant co-moving entropy density $\sigma$
and  the relation $Ts=\rho+p$ for the physical entropy density $s$.
It follows that either the entropy or the temperature is
negative for the phantom. Applying the constant co-moving entropy density
to the first law of thermodynamics for DE with constant equation of state
$w=p/\rho$, the thermodynamical properties of DE and the thermal spectral
distribution were discussed in \cite{lima}. The thermodynamics of the
DE with constant $w$ in the range $-1<w-1/3$ was also
discussed in \cite{bousso05}, and the thermodynamics of phantom
with $w<-1$ was investigated in \cite{pedro}. Some other discussions on this
topic can be found in \cite{odintsov,santos}.
However, in general, when we consider the thermal equilibrium state of the Universe,
the temperature of the Universe is associated with the
horizon, then the integrability condition cannot be used to derive the constancy of the
co-moving entropy density from the first law of thermodynamics. Some
assumptions on the temperature or entropy are needed to derive the thermodynamical
properties of DE. With the requirement that the entropy of
DE is bounded from above and the assumption that it
could not increase faster than that of cosmic microwave background radiation,
we will investigate the thermodynamical properties of
DE in the thermal equilibrium universe.
Employing the holographic entropy bound, we will also
discuss the GSL for the Universe with DE.
The study of GSL in a phantom
dominated universe was carried out in \cite{phanthermo}.
We will address the thermodynamics of DE by considering
the DE models with constant $w$ and the
generalized Chaplygin gas (GCG) model
which extrapolates between a dust and a cosmological constant \cite{chap}.

The paper is organized as follows. In section \ref{aphor}, we review the equivalence
between Friedmann equation and the first law of thermodynamics of the apparent
horizon. In section \ref{vtind}, we discuss the thermodynamical
consequence by assuming the physical volume and temperature are independent. In section \ref{vtdep},
considering the physical volume is a function of the temperature,
we investigate the thermodynamical properties for the DE with constant $w$ and the generalized
Chaplygin gas. We discuss the GSL of thermodynamics for the Universe with DE in section \ref{gsl}.
The summary and discussion will be presented in the last section \ref{conc}.

\section{First law of thermodynamics on the apparent horizon}
\lb{aphor}

For a spherically symmetric space-time with the metric
$ds^2=g_{ab}dx^a dx^b+\tilde{r}^2 d\Omega^2$, where the unit
spherical metric $d\Omega^2=d\theta^2+\sin^2\theta d\phi^2$,
we define the Misner-Sharp mass $M$ as \cite{israel,misner}
\bq
\lb{mfunc}
g^{ab}\tilde{r}_{,a}\tilde{r}_{,b} \equiv f \equiv
1-\frac{2GM}{\tilde{r}}.
\eq
The mass $M$ defined in Eq.
(\ref{mfunc}) was shown to be the active gravitational energy in the
vacuum, small sphere, large sphere, Newtonian, test particle and
relativistic limits \cite{sean}.
The Einstein field equations tell us \cite{israel}
\bq
\lb{masseq}
M_{;a}=4\pi\tilde{r}^2(T_a^b-\delta_a^b T^c_c)\tilde{r}_{;b},
\eq
where the semicolon denotes covariant derivative with respect to
the two dimensional metric $g_{ab}$.
The apparent horizon is determined from $f=0$.
We define the dynamic surface gravity at the apparent horizon as
\bq
\lb{kappa}
\kappa\equiv
-\frac{1}{2}\partial_{\tilde{r}} f\Big|_{f=0}.
\eq
So the Hawking temperature at the apparent horizon is $T=|\kappa|/2\pi$.
The surface gravity $\kappa$ defined in Eq. (\ref{kappa}) has the
same form as the Newtonian surface gravity. In static black hole physics,
the horizon is located at $f(\tilde{r})=1-2GM/\tilde{r}=0$ which
gives $\tilde{r}=2GM$. The surface gravity is found to be
$\kappa=-f'/2=-1/4GM$ and the Hawking temperature is $T=1/8\pi G M$.
For the de Sitter universe, $f(\tilde{r})=1-H^2\tilde{r}^2$,
the event horizon which coincides with the apparent horizon
is $\tilde{r}_A=H^{-1}$.
The surface gravity at the apparent (event) horizon is found to be
$\kappa=-f'/2=\tilde{r}^{-1}_A$ and the Hawking temperature
$T=1/2\pi \tilde{r}_A$. For static spherically symmetric space-time,
it was shown that Einstein equation at the apparent horizon $\tilde{r}_i$ gives the first law of
thermodynamics of the apparent horizon with the identification of the temperature as
$T=|f'(\tilde{r}_i)|/4\pi$ and the entropy as $S=A/4G$, where $A=4\pi \tilde{r}_i^2$ is the
area of the horizon \cite{padmanabhan02}. It was shown that the apparent
horizon is a good boundary holding thermodynamical laws
in the Universe driven by quintessence and phantom \cite{wang06}.

For the Friedmann-Robertson-Walker (FRW) metric, we have
\bq
\lb{frwmetric}
g_{ab}=\begin{pmatrix}
  -1 & 0 \\
0 & a^2/(1-kr^2)
\end{pmatrix}, \quad \tilde{r}=ar.
\eq
So
\bq
\lb{fval1}
f=g^{ab}\tilde{r}_{,a}\tilde{r}_{,b}=1-\left(H^2+\frac{k}{a^2}\right)\tilde{r}^2.
\eq
The apparent horizon is
\bq
\lb{apphor}
\tilde{r}_A=ar_A=\frac{1}{\sqrt{H^2+k/a^2}}.
\eq
The surface gravity at the apparent horizon is $\kappa=-f'/2=1/\tilde{r}_A$
and the Hawking temperature is
\bq
\lb{tapparent}
T_A=\frac{1}{2\pi \tilde{r}_A}.
\eq

By using the FRW metric (\ref{frwmetric}), the definition of the Misner-Sharp mass
(\ref{mfunc}), and the perfect fluid plus the cosmological constant for the matter,
we get the Friedmann equations
from the mass formulae (\ref{masseq}),
\begin{gather}
\lb{freq1}
H^2+\frac{k}{a^2}=\frac{8\pi G}{3}\rho+\Lambda,\\
\lb{freq2}
\dot{\rho}+3H(\rho+p)=0,\\
\lb{freq3}
\dot{H}-\frac{k}{a^2}=-4\pi G(\rho+p).
\end{gather}
Eq. (\ref{freq3}) can be derived from Eqs. (\ref{freq1}) and (\ref{freq2}), so there are only
two independent equations.

Combining Eqs. (\ref{apphor}) and (\ref{freq3}), we have
\bq
\lb{rateq}
\dot{\tilde{r}}_A=-\tilde{r}_A^3 H(\dot{H}-k/a^2)=4\pi G \tilde{r}_A^3 H(\rho+p).
\eq
The energy flow through the apparent horizon is
\bq
\lb{eflow1}
-dE=4\pi \tilde{r}^2_A T_{\mu\nu}k^\mu k^\nu dt=4\pi\tilde{r}^3_A(\rho+p)H dt,
\eq
where the (approximate) Killing vector or the (approximate) generators
of the horizon, the future directed
ingoing null vector field $k^\mu=(1,\ -Hr,\ 0,\ 0)$.
Note that $dE=k^a M_{;a}$ with $k^a\tilde{r}_{;a}=0$.
Therefore, we get the first law of thermodynamics
of the apparent horizon
\bq
\lb{slaw1}
T_A dS_A=\frac{1}{2\pi \tilde{r}_A}\frac{d}{dt}\left(\frac{4\pi \tilde{r}_A^2}
{4G}\right)=\dot{\tilde{r}}_A/G dt=-dE,
\eq
if we define the entropy of the apparent horizon as $S_A=A/4G=\pi \tilde{r}_A^2/G$.
It is clear that the first law of thermodynamics can be derived from Einstein equations.
More interestingly, Einstein equations can be obtained from the first law.
By using the first law of thermodynamics, we get
\bq
\lb{radot}
\dot{\tilde{r}}_A=4\pi G\tilde{r}_A^3 H(\rho+p)=-\tilde{r}_A^3 H(\dot{H}-k/a^2).
\eq
So
\bq
\lb{freq4}
\dot{H}-k/a^2=-4\pi G(\rho+p).
\eq
Combining Eq. (\ref{freq4}) with the energy conservation equation (\ref{freq2}),
we can derive Friedmann equation (\ref{freq1}).
In this derivation, there is also an integration constant which is set to be the cosmological constant.
The derivation of Friedmann equation from thermodynamics was also shown for gravity with
Gauss-Bonnet term and Lovelock gravity theories \cite{cai06}.
The equivalence between the Friedmann equation and the first law of thermodynamics
of the apparent horizon for other gravitational theories, like the higher
order gravity $f(R)$ and the Brans-Dicke theory, remains to be an open problem.

There are other general discussions on the first law of thermodynamics.
In \cite{sean}, the author derived a so called unified first law from the mass
formulae (\ref{mfunc}),
\bq
\lb{masseq1}
\nabla_a M= A \Psi_a +W \nabla_a V,
\eq
where the energy density
$W=-\frac{1}{2}T^a_a$,
the energy-supply vector
$\Psi_a=T_a^b\partial_b\tilde{r}+W\partial_a\tilde{r}$,
and the areal volume is determined by $\nabla V=A\nabla \tilde{r}$.
The first term $A\Psi$ in the unified first law may
be interpreted as an energy-supply term, analogous to the heat-supply term in
the classical first law of thermodynamics, while the second term $W\nabla V$ may
be interpreted as a work term. If we define the variation along the trapping horizon
as $\delta M=z^a\nabla_a M$, where $z$ is a vector tangent to the trapping horizon
and the normalization of $z$ is irrelevant, then it was shown that
$Az^a\Psi_a=\kappa_1 \delta A/8\pi$ and the unified first law becomes
\bq
\lb{seanslaw}
\delta M=\frac{\kappa_1}{8\pi}\delta A+W\delta V,
\eq
where $\kappa_1=\Box\tilde{r}/2=M/\tilde{r}^2-4\pi\tilde{r} W$. For
cosmological models with FRW metric, $\kappa_1\neq \kappa$ and the above
first law (\ref{seanslaw}) is different from the one (\ref{slaw1})
derived from the Friedmann equation directly. Although the
unified first law seems to be true for more general geometry with
trapping horizon,
our definitions of the surface gravity and the Hawking
temperature (\ref{kappa}) are appropriate for use
in the cosmological context since the first
law built reflects the requirement of the Einstein field equation.

Since at the apparent horizon, $f=0$, so the vector $\nabla_a f$ is normal to
the apparent horizon, while the orthogonal vector $\zeta^a=\epsilon^{ab}\nabla_b f$
is tangent to it. If we define the variation at the apparent horizon as
$\delta M=\zeta^a \nabla_a M$, then we have \cite{frolov}
\bq
\lb{fslaw}
\delta E=\zeta^a M_{;a}=\frac{M}{\tilde{r}_A^2}\tilde{r}_A
\zeta^a\tilde{r}_{A;a}
=\frac{1}{4\pi G\tilde{r}_A}\zeta^a(\pi\tilde{r}_A^2)_{;a}
=\frac{T_A}{2}\delta S.
\eq
In the above derivation, only the definitions of $M$ and $\tilde{r}_A$
are used. Although the formulae (\ref{fslaw}) looks in form like the first law
of thermodynamics, it is different from Eq. (\ref{slaw1}).
The physical meanings of $\delta E$ and $\delta S$ are
different from $dE$ and $dS$ before. Further, Eq.(\ref{fslaw}) cannot
reflect the dynamical property of the Universe.

Thus, in the cosmological context, the apparent horizon
is the physical proper size in discussing thermodynamics and
the appropriate description of the first
law on the apparent horizon is expressed in Eq. (\ref{slaw1}) with the definitions of
the temperature in Eq. (\ref{tapparent})
and the entropy $S=\pi \tilde{r}^2_A/G$ on the apparent horizon.
The first law for more general spherically symmetric space-times in other
gravity theories needs further investigation.

\section{Constant co-moving entropy density}
\lb{vtind}
In an expanding universe,
the first law of thermodynamics is
\bq
\label{2nd}
TdS(T,V)=Td(s V)=d(\rho(T)V)+p(T)dV,
\eq
where
$s$ is the entropy density and $V$ is the physical volume. The energy density and the
pressure are assumed to be functions of temperature only,
and the volume $V$ and the temperature $T$ are usually assumed to be independent.
The integrability condition $\partial^2 S/\partial V\partial
T=\partial^2 S/\partial T\partial V$ relates the energy density and pressure
\bq
\label{ptrel}
\frac{dp}{dT}=\frac{\rho+p}{T}.
\eq
For the cosmological constant $p=-\rho$, the first law
implies $dS=d[V(\rho+p)/T]=0$. Substitute Eq. (\ref{ptrel}) into Eq.
(\ref{2nd}), we get
\bq
\label{ptrel1}
s=\frac{\rho+p}{T},\
{\rm for}\ \rho\neq -p.
\eq
For phantom field, we see that
either the temperature or the entropy is negative. Combining Eqs.
(\ref{2nd}), (\ref{ptrel}) and (\ref{ptrel1}), we get
\bq
\label{ptrel2}
ds=\frac{d\rho}{T}=s
\frac{d\rho}{\rho+p},\ {\rm for}\ \rho\neq -p.
\eq
Therefore, once
we are given the equation of state $p/\rho$, we can get the
relationship between the entropy density $s$ and the energy
density $\rho$, thence we can find the relationship between the
temperature and the energy density. If we substitute the energy
conservation equation (\ref{freq2}) into Eq. (\ref{ptrel2}), we get
\bq
\label{rel4}
\sigma=s a^3={\rm constant}.
\eq
Therefore, the first law of thermodynamics combined with
energy conservation tell us that the co-moving entropy density $\sigma$
is a constant. Note that the above
results (\ref{ptrel1})-(\ref{rel4}) are independent of the volume.
They are derived based on the following assumptions: (1) The energy density
and pressure are functions of temperature only; (2) The entropy
is extensive so that we can define entropy density; (3) The volume
and the temperature are independent.

In the discussion of the phantom thermodynamics, it was assumed that
the temperature of DE is proportional to
the horizon temperature, i.e., $T=bT_A$ \cite{phanthermo}. For $b=1$,
the DE is in thermal equilibrium with the horizon. We adopt this temperature of the Universe $T=bT_A$ to
discuss the DE thermodynamical property. Substituting $T$
into Eq. (\ref{ptrel1}),  using
Eqs. (\ref{freq3}) and (\ref{rel4}), we get
\bq
\lb{asol1}
s=\frac{\sigma}{a^3}=-\frac{\dot{H}-k/a^2}{2bG(H^2+k/a^2)^{1/2}}.
\eq
Considering
the Friedmann equation (\ref{freq3}), we see that for quintessence
with $w>-1$, $s>0$, while for phantom with $w<-1, s<0$.
For the flat universe
$k=0$, the solution to Eq. (\ref{asol1}) is
\bq
\lb{asol2}
a(t)=[2b\sigma G(t-t_*)]^{1/3},
\eq
where $t_*$ is an integration constant, or
\bq
\lb{asol3}
a(t)=\left[e^{3d(t-t_*)}-\frac{2b\sigma G}{3d}\right]^{1/3},
\eq
where $d\neq 0$ is an integration constant which is the asymptotical Hubble parameter.
Only Eq. (\ref{asol3}) gives the acceleration of the Universe driven by the DE.
If $\sigma=0$, Eq. (\ref{asol3}) becomes the cosmological constant solution.
Substituting Eq. (\ref{asol3}) into Friedmann equations
(\ref{freq1})-(\ref{freq3}), we get the DE equation of state
\bq
\lb{desteq}
p=-\rho\left[1-\frac{4b \sigma G}{3d}e^{-3d(t-t_*)}\right].
\eq
Eq. (\ref{desteq}) is the DE equation of state when the DE temperature
$T=bT_A$. The DE is in thermal equilibrium with the apparent horizon when $b=1$.
We see that when the DE is the cosmological constant, $\sigma=0$ and $p=-\rho$.
If the DE persists with an equation of state different from that described in Eq. (\ref{desteq}), it
concludes that DE is not in thermal equilibrium with the apparent horizon in general.

\subsection{Dark energy with constant $w$}
\label{fixw}

For a fluid with constant equation of state $p=w\rho$, the
solution to the energy conservation equation (\ref{freq2}) is
$\rho_w=\rho_{w0}(a_0/a)^{3(1+w)}$.
From Eq. (\ref{ptrel2}) we learn that
\bq
\lb{ws1}
\frac{s_w}{s_{w0}}=
\left(\frac{\rho_w}{\rho_{w0}}\right)^{1/(1+w)}
=\left(\frac{a}{a_0}\right)^{-3},
\eq
where the subscript ``0" denotes the present time value.
The entropy inside the apparent horizon is
\bq
\lb{sweq1}
S_w=\frac{4}{3}\pi s_w \tilde{r}_A^3=S_{w0}
\left(\frac{a}{a_0}\right)^{-3}
\left(\frac{\rho_t}{\rho_{t0}}\right)^{-3/2},
\eq
where $\rho_t$ is the total energy density and the Friedmann equation
(\ref{freq1}) was used to derive the last equation. During the radiation
domination (RD), $\rho_t=\rho_r\sim a^{-4}$,
$S_R=S_{R0}\Omega_{r0}^{-3/2}(a/a_0)^3$ and
$S_A=\pi\tilde{r}^2_A/G=S_{A0}\Omega_{r0}^{-1}(a/a_0)^4$, where
$\Omega_{r0}=\rho_{r0}/\rho_{t0}$. During the matter domination (MD), $\rho_t=\rho_m\sim a^{-3}$,
so $S_R=S_{R0}\Omega_{m0}^{-3/2}(a/a_0)^{3/2}$
and $S_A=S_{A0}\Omega_{m0}^{-1}(a/a_0)^3$. During the $w$-fluid
domination, $\rho_t=\rho_w\sim a^{-(1+w)}$, then $S_R=S_{R0}\Omega_{w0}^{-3/2}
(a/a_0)^{3(1+3w)/2}$ and $S_A=S_{A0}\Omega_{w0}^{-1}(a/a_0)^{3(1+w)}$.
During RD and MD eras, the entropy increases as the Universe expands, but when the DE dominates $w<-1/3$,
 the total entropy decreases as the Universe expands.
It is easy to check that the entropy in each epoch of the evolution of the Universe
is bounded by the apparent horizon entropy, once
the entropy bound is satisfied at an earlier era.
At very early epoch, the entropy bound is
\bq
\lb{abound}
\frac{S_R}{S_A}=\frac{S_{R0}}{S_{A0}}\Omega_{r0}^{-1/2}
\left(\frac{a_0}{a}\right)\le 1.
\eq
So $a/a_0=T_{\gamma 0}/T_{\gamma} \ge
S_{R0}\Omega_{r0}^{-1/2}/S_{A0}\sim 10^{-32}$ if we choose
$H_0=72$ Mpc$^{-1}$ km s$^{-1}$, $S_{R0}
\sim  10^{88}$, $\Omega_{r0}\sim 10^{-4}$ and $S_{A0}\sim 10^{122}$,
thus we have $T_\gamma\lesssim 10^{19}$GeV.
The relationship (\ref{ws1}) between $s$ and $\rho$ may be used
to provide other useful information, for example, if there is an
energy bound, then we can get an entropy bound, which was used to argue the
minimum mass of a primordial black hole in \cite{bkmass}.

Substitute Eq.
(\ref{ws1}) into Eq. (\ref{ptrel1}), we get
\bq
\lb{wt1}
\frac{T_w}{T_{w0}}=\left(\frac{\rho_w}{\rho_{w0}}\right)^{w/(1+w)}
=\left(\frac{a}{a_0}\right)^{-3w},
\eq
where $T_{w0}=(1+w)\rho_{w0}/s_{w0}$. For the radiation,
$w=1/3$, we get the familiar relation $\rho\sim T^4$.
For the DE, we see that it becomes hotter and hotter as
the Universe expands. At present, the DE density $\rho_{w0}$ is much
greater than that of the radiation $\rho_{r0}$, the DE
temperature could be much higher than the current cosmic microwave
background temperature unless $w$ is very close to $-1$.

In \cite{lima}, the authors find the spectral distribution
$E^{1/w}/[\exp(E/T)\pm 1]$. Assuming that the DE is massless
and follows either the Bose-Einstein or Fermi-Dirac statistics,
we get
\bq
\lb{wspectrum}
\rho=\frac{g}{2\pi^2}\int_0^\infty\frac{E(P)}{\exp[E(P)/T]\pm
1}P^2dP \sim \frac{g}{6\pi^2
w}\int_0^\infty\frac{E^{1/w}}{\exp[E(P)/T]\pm
1}dE,
\eq
where $P$ is
the momentum and $g$ is the internal degrees of freedom.  The
number density is
\bq
\lb{wnumber}
n=\frac{g}{2\pi^2}\int_0^\infty\frac{P^2 dP}{\exp(E/T)\pm
1}\sim \frac{g T^{1/w}}{6\pi^2 w}\int_0^\infty
\frac{u^{1/w-1}du}{e^u\pm 1}=n_0\left(\frac{T}{T_0}\right)^{1/w}.
\eq
Therefore, we get the dispersion relationship $E(P)\sim P^{3w}$.
For radiation, we recover the usual dispersion relationship $E=P$.
For DE, we see that the greater the momentum, the
smaller the energy. This is another strange property of DE.
It could be that we cannot apply the above
reasoning to the DE, i.e., the DE may not follow
the usual statistics or it is not massless.

\subsection{The generalized Chaplygin gas}
\label{gcg}

For the GCG model, $p_c=-A/\rho_c^\alpha$ $(\alpha>0)$, Eqs.
(\ref{ptrel1}) and (\ref{ptrel2}) give us that
\begin{gather}
\lb{chapsd}
s^{\alpha+1}_c=C(\rho_c^{\alpha+1}-A),\\
\lb{chapte}
T=C^{-1/(1+\alpha)}\left(1-\frac{A}{\rho_c^{\alpha+1}}\right)^{\alpha/\alpha+1}.
\end{gather}
The above relationships were derived in \cite{santos}. In terms of
the variables $w_{c0}$ and $\rho_{c0}$, we can write
$A=-w_{c0}\rho_{c0}^{\alpha+1}$. Eqs. (\ref{chapsd}) and
(\ref{chapte}) can be re-written as
\begin{gather}
\lb{chapsde}
\left(\frac{s_c}{s_{c0}}\right)^{\alpha+1}=\frac{1}{1+w_{c0}}
\left[\left(\frac{\rho_c}{\rho_{c0}}\right)^{1+\alpha}+w_{c0}\right],
\\
\lb{chapet}
\rho_c=\rho_{c0}\left[-\frac{1}{w_{c0}}+\frac{1+w_{c0}}{w_{c0}}
\left(\frac{T}{T_0}\right)^{(1+\alpha)/\alpha}\right]^{-1/(1+\alpha)},
\end{gather}
where
$T_{c0}=(1+w_{c0})^{\alpha/(1+\alpha)}C^{-1/(\alpha+1)}
=(1+w_{c0})\rho_{c0}/s_{c0}$.

The solution to the energy conservation equation (\ref{freq2}) is
\bq
\lb{chapden}
\rho_c=\rho_{c0}\left[-w_{c0}+(1+w_{c0})\left(\frac{a_0}{a}
\right)^{3(1+\alpha)}\right]^{1/(1+\alpha)}.
\eq
This equation tells us that the GCG energy density
decreases as the Universe expands and $\rho_c\rightarrow
\rho_{c0}(-w_{c0})^{1/(1+\alpha)}$ as $a\rightarrow \infty$, i.e.,
the GCG extrapolates between a dust and a cosmological
constant. From Eq. (\ref{chapet}), we see that the temperature of
the GCG starts from the maximum temperature $T_m=T_0
(1+w_{c0})^{-\alpha/(1+\alpha)}$ and decreases to zero as the
Universe expands. The temperature behavior of the GCG is
different from that of the DE with constant $w$. The
temperature of the GCG decreases instead of increasing.
This tells us that the increase or decrease behavior of the temperature of the Universe dominated by DE
is model dependent, it is not the general property associated with the DE.

\section{Volume and temperature inside the apparent horizon}
\lb{vtdep}

In cosmology, the physical volume is a function of the scale factor $a(t)$. Since the
scale factor $a(t)$ relates to the temperature $T$ of the Universe, the volume $V$ should be
a function of $T$. The first law (\ref{2nd}) becomes
\bq
\lb{theq1}
(Ts-\rho-p)dV=V(d\rho-Tds).
\eq
The formula (\ref{ptrel1})-(\ref{rel4}) are special solution
to Eq. (\ref{theq1}) and they are not valid
in general. As discussed in section \ref{aphor},
we should use the apparent horizon in the discussion of
thermodynamics in cosmology \cite{wang06}.
For the radiation entropy, the previous results still
hold. During RD, the entropy contributed by relativistic particles is
$S_R/S_{R0}=\Omega_{r0}^{-3/2}(a/a_0)^3$.
During MD, $S_R/S_{R0}=\Omega_{m0}^{-3/2}(a/a_0)^{3/2}$.
Substituting $V=4\pi \tilde{r}_A^3/3$ into
Eq. (\ref{theq1}) and using Eq. (\ref{freq1}), we get
\bq
\lb{theq2}
\begin{split}
TdS&=-\frac{2\pi}{3}\left(\frac{8\pi G}{3}\right)^{-3/2}\rho_t^{-5/2}
[3(\rho+p)d\rho_t-2\rho_td\rho]\\
&=-\frac{2\pi}{3}\left(\frac{8\pi G}{3}\right)^{-3/2}\rho_t^{-5/2}
(\rho_t+3p_t)d\rho,
\end{split}
\eq
where the second equality was obtained by using the energy conservation law.
During DE domination, $TdS>0$ if $p<-\rho$ and
$TdS\le 0$ if $p\ge -\rho$.

\subsection{Dark energy in thermal equilibrium}

If we think that DE is in thermal equilibrium with
the Hawking radiation of the apparent horizon, i.e.,
$T=T_A=1/2\pi\tilde{r}_A$, then we get
\bq
\lb{theq3}
dS=-\frac{3}{16G^2}(\rho_t+3p_t)\rho_t^{-3}d\rho.
\eq

For dark energy $p_w=w\rho_w$, the solution to Eq. (\ref{theq3})
during RD is
\bq
\lb{swrd1}
S_w=\frac{3(1+w)}{5-3w}S_{A0}\Omega_{r0}^{-2}\Omega_{w0}
\left(\frac{a}{a_0}\right)^{5-3w}.
\eq
This is the DE entropy during RD period.
During MD, the DE entropy can be obtained from Eq. (\ref{theq3}) in the form
\bq
\lb{swmd1}
S_w=S_{w1}+\frac{1+w}{2(1-w)}S_{A0}\Omega_{m0}^{-2}\Omega_{w0}
\left(\frac{a}{a_0}\right)^{3(1-w)},
\eq
where $S_{w1}$ is an integration constant.
We see that the entropy of the DE increases faster than
those of the radiation and the apparent horizon during RD and MD.
To require that the current
DE entropy is negligible compared to the radiation entropy, we get
\bq
\lb{weq1}
\frac{1+w}{2(1-w)}\left(\frac{a_t}{a_0}\right)^{3/2-3w}\ll
\frac{S_{R0}\Omega_{m0}^{1/2}}{S_{A0}\Omega_{w0}}
\sim 10^{-34},
\eq
where $a_t$ is taken to be the end of the MD which is roughly the
transition from deceleration to acceleration. The requirement of neglecting the DE entropy
puts the constraint on the DE equation of state to be very close to $-1$. This is consistent with
recent observations. If the DE equation of state is different from $-1$, then the DE entropy
dominates the entropy components when the DE is in the thermal equilibrium
with the apparent horizon. Comparing the DE entropy with the apparent horizon entropy
\bq
\lb{wsbound1}
\frac{S_w}{S_A}\Big|_{a\simeq a_t}=\frac{1+w}{2(1-w)}
\frac{\Omega_{w0}}{\Omega_{m0}}\left(\frac{a_t}{a_0}\right)^{-3w}<1,
\eq
we find that the holographic bound does not put strict constraint on $w$ and it
can be satisfied for general DE equation of state.

\begin{figure}[htp]
\centering
\includegraphics[width=12cm]{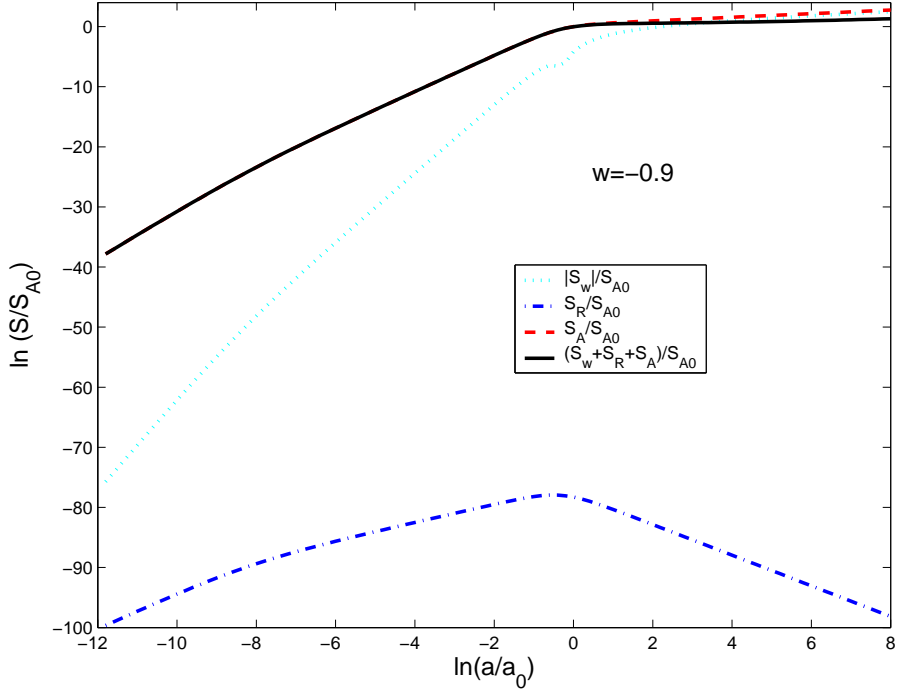}
\caption{The evolutions of $S_w$, $S_R$ and $S_A$. The dotted line is for
$|S_w|/S_{A0}$ (note that $S_w$ is negative after the MD), the
dash-dot line is for $S_R/S_{A0}$, the dashed line is for $S_A/S_{A0}$,
and the solid line is for the total entropy $(S_A+S_R+S_w)/S_{A0}$.}
\label{fig1}
\end{figure}

During DE domination, the radiation entropy contributed by relativistic particles becomes
$S_R/S_{R0}=\Omega_{w0}^{-3/2}(a/a_0)^{3(1+3w)/2}$, the
apparent horizon entropy becomes $S_A/S_{A0}=\Omega_{w0}^{-1}(a/a_0)^{3(1+w)}$
and the solution to Eq. (\ref{theq3}) gives the DE entropy
\bq
\lb{wseq4}
S_w=S_{wi}+\frac{3}{16 G^2}(1+3w)\rho_w^{-1}
=S_{wi}+\frac{1+3w}{2}S_{A0}\Omega_{w0}^{-1}\left(\frac{a}{a_0}\right)^{3(1+w)},
\eq
where $S_{wi}$ is an integration constant.
For DE with $w<-1/3$, the radiation entropy decreases while both the apparent horizon
entropy and the absolute value of the DE entropy increase at the same
rate. Eventually the radiation entropy will vanish and the DE
entropy will become negative whose statistical meaning becomes hard to understand.
The total entropy of the space-time is
$S_t=S_w+S_A=S_{wi}+3(1+w)S_{A0}\Omega_{w0}^{-1}(a/a_0)^{3(1+w)}/2$.
Therefore the total
entropy is positive if $w\ge -1$. To see the above results more clearly,
we solve Eq. (\ref{theq3}) numerically for $w=-0.9$ and the result is
shown in Fig. \ref{fig1}. It is evident that $S_w$ increases faster than
$S_R$ and $S_w>S_R$ during RD and MD, and $S_w$ is negative after MD. The
entropy bound is always satisfied.

For the GCG, Eq. (\ref{chapden}) tells us that
$\rho_c=\rho_{c0}(1+w_{c0})^{1/(1+\alpha)}(a_0/a)^3$ during RD and
the solution to Eq. (\ref{theq3}) is
\bq
\lb{chaps1}
S_c=\frac{3}{5}S_{A0}\Omega_{r0}^{-2}(1+w_{c0})^{1/(1+\alpha)}\left(
\frac{a}{a_0}\right)^5.
\eq
The entropy of the GCG increases faster than that of the radiation.

\begin{figure}[htp]
\centering
\includegraphics[width=12cm]{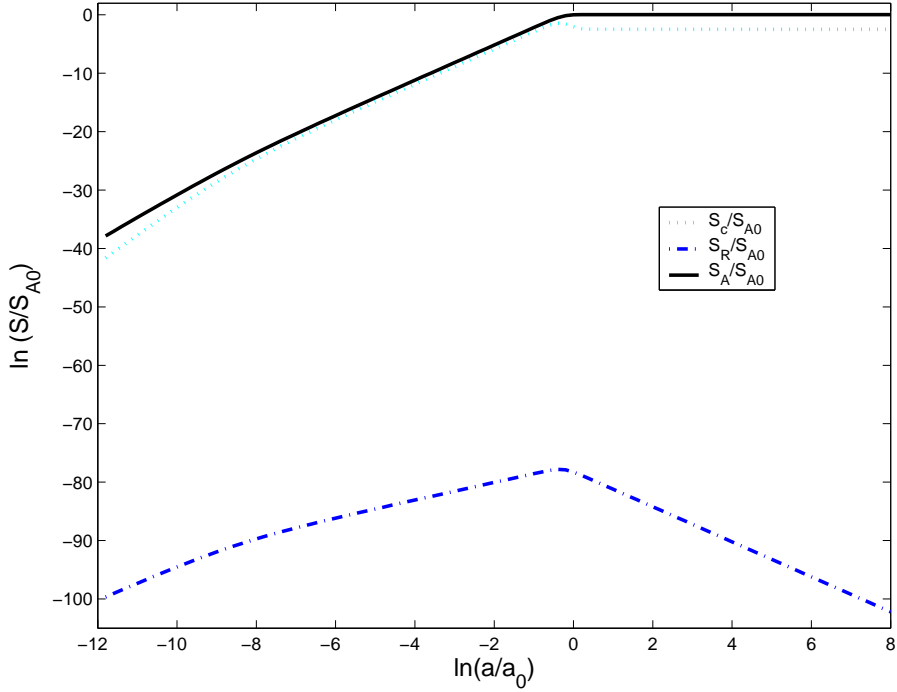}
\caption{The evolutions of $S_c$, $S_R$ and $S_A$ with $w_{c0}=-0.88$ and $\alpha=1.57$.
The dotted line is for $S_c/S_{A0}$, the
dash-dot line is for $S_R/S_{A0}$, and the solid line is for $S_A/S_{A0}$.}
\label{fig2}
\end{figure}

During the epoch of the GCG domination,
the solution to Eq. (\ref{theq3}) is
\bq
\lb{chapsol4}
\begin{split}
S_c&=S_{ci}+\frac{1}{2}S_{A0}\left(\frac{\rho_{c0}}{\rho_c}\right)^{\alpha+2}
\left[\left(\frac{\rho_c}{\rho_{c0}}\right)^{\alpha+1}+
\frac{3w_{c0}}{\alpha+2}\right]\\
&=\begin{cases}
S_{ci}+\frac{1}{2}S_{A0}(1+w_{c0})^{-1/(1+\alpha)}(a/a_0)^3,&
a\ll a_0,\\
\frac{\alpha-1}{2(2+\alpha)} (-w_{c0})^{-1/(1+\alpha)} S_{A0},& a\gg a_0,
\end{cases}
\end{split}
\eq
where $S_{ci}$ is an integration constant.
It is easy to see that $S_{c0}=S_{A0}(\alpha+2+3w_{c0})/2(\alpha+2)$ and
the holographic entropy bound is satisfied.
If $\alpha\ge 1$, then the entropy is always non-negative and it increases first and
then decreases. The apparent horizon entropy is
\bq
\lb{sachap1}
\begin{split}
S_A&=S_{A0}\left[-w_{c0}+(1+w_{c0})\left(\frac{a}{a_0}\right)^{-3(1+\alpha)}
\right]^{-1/(1+\alpha)}\\
&=\begin{cases}
(1+w_{c0})^{-1/(1+\alpha)}S_{A0} (a/a_0)^3,& a\ll a_0,\\
(-w_{c0})^{-1/(1+\alpha)} S_{A0},& a\gg a_0.
\end{cases}
\end{split}
\eq
The entropy of the GCG is always
smaller than that of the apparent horizon
and the holographic bound is always respected. The radiation entropy is
\bq
\lb{chaprs1}
\begin{split}
S_R&=S_{R0}\left(\frac{a}{a_0}\right)^{-3}\left[-w_{c0}
+(1+w_{c0})\left(\frac{a}{a_0}\right)^{-3(1+\alpha)}\right]^{-3/2(1+\alpha)}\\
&=\begin{cases}
(1+w_{c0})^{-3/2(1+\alpha)}S_{r0}(a/a_0)^{3/2},& a\ll a_0,\\
 (-w_{c0})^{-3/2(1+\alpha)} S_{r0}(a/a_0)^{-3},
& a\gg a_0.
\end{cases}
\end{split}
\eq
The radiation entropy increases first and then decreases to zero. Since the
entropy of the GCG increases faster than that of the radiation,
the GCG entropy is greater than the radiation entropy at
the present because $S_{c0}>S_{r0}$ and the holographic entropy bound is always respected.
These results are evident from Fig. \ref{fig2} with $w_{c0}=-0.88$ and $\alpha=1.57$ \cite{chap}.
So the GCG entropy is the dominate entropy if it is in thermal equilibrium with the apparent horizon.

\subsection{Dark energy with power law entropy}
\label{power}

Recall that the radiation temperature $T\sim a^{-1}$, the radiation entropy
$S\sim T^{-3}$ during RD and $S\sim T^{-3/2}$ during MD. The relationship
between the entropy and the temperature is a power law form.
More generally, we assume that $S=\beta T^\gamma$, where the constants
$\beta$ and $\gamma$ depend on the background,
then Eq. (\ref{theq2}) becomes
\bq
\lb{theq5}
\gamma\beta T^\gamma dT=-\frac{2\pi}{3}
\left(\frac{8\pi G}{3}\right)^{-3/2}(\rho_t+3p_t)\rho_t^{-5/2}d\rho.
\eq

For the DE with $p=w\rho$,
the solution to Eq. (\ref{theq5}) during RD is
\bq
\lb{rdsw5}
\frac{\gamma_{w1}\beta_{w1}}{\gamma_{w1}+1}T_w^{\gamma_{w1}+1}
=\frac{4\pi(1+w)}{3(1-w)}
\left(\frac{8\pi G}{3}\right)^{-3/2}\rho_{r0}^{-3/2}\rho_{w0}
\left(\frac{a}{a_0}\right)^{3(1-w)}.
\eq
For radiation, $\gamma_{w1}=-3$ and $T_w\sim a^{-1}$.
If $\gamma_{w1}<-1$ for $-1/3>w\ge -1$, then the DE temperature
decreases and the entropy increases as the Universe expands. Since
$S_R\sim a^3$ and $S_w\sim a^{3(1-w)\gamma_{w1}/(1+\gamma_{w1})}>a^3$, the
DE entropy increases faster than the radiation entropy. To make
sure that the DE entropy is negligible compared to the radiation
entropy, we require $0<\gamma_{w1}<-1/w$ so that $\gamma_{w1}/(1+\gamma_{w1})>0$
and $3(1-w)\gamma_{w1}/(1+\gamma_{w1})<3$ for $-1/3>w\ge -1$. Therefore the DE
temperature and entropy both increase as the Universe expands.
For $w<-1$, we need $-1<\gamma_{w1}<0$ to keep the temperature positive.
The phantom temperature increases and the
entropy decreases as the Universe expands. However, both the temperature
and the entropy of the phantom are positive.

During MD, the solution to Eq. (\ref{theq5}) is
\bq
\lb{mdsw5}
\frac{\gamma_{w2}\beta_{w2}}{\gamma_{w2}+1}T_w^{\gamma_{w2}+1}
=C_1+\frac{4\pi(1+w)}{3(1-2w)}
\left(\frac{8\pi G}{3}\right)^{-3/2}\rho_{m0}^{-3/2}\rho_{w0}
\left(\frac{a}{a_0}\right)^{3(1-2w)/2},
\eq
where $C_1$ is an integration constant.
If $\gamma_{w2}<-1$ for $-1/3>w\ge -1$, then the DE entropy
increases at a faster rate than
that of the radiation as the Universe expands.
So we require $0<\gamma_{w2}<-1/2w$ to keep $\gamma_{w2}/(1+\gamma_{w2})>0$
and $3(1-2w)\gamma_{w2}/2(1+\gamma_{w2})<3/2$. Again both the temperature
and entropy of the DE with $w\ge -1$ increase as the Universe expands.
For $w<-1$, we require $-1<\gamma_{w2}<0$. The phantom temperature
increase and the entropy decreases as the Universe expands.

During the DE domination,
the solution to Eq. (\ref{theq5}) is
\bq
\lb{soleq5}
\frac{\gamma_{w3}\beta_{w3}}{\gamma_{w3}+1}T_w^{\gamma_{w3}+1}=C_2+\frac{4\pi}{3}
\left(\frac{8\pi G}{3}\right)^{-3/2}(1+3w)
\left(\frac{a}{a_0}\right)^{3(1+w)/2},
\eq
where $C_2$ is an integration constant.  Now we require that
$-1<\gamma_{w3}< 0$ for both the quintessence and phantom
to get a positive temperature.
For $w\ge -1$, the temperature increases and the entropy decreases
as the Universe expands. For $w<-1$, the temperature
decreases and the entropy increases as the Universe expands.
At late times, the entropy is
\bq
\lb{wseq5}
\begin{split}
S_w&=\beta_{w3}\left[\frac{4\pi}{3\beta_{w3}}\left(
\frac{8\pi G}{3}\right)^{-3/2}(1+3w)
\frac{\gamma_{w3}+1}{\gamma_{w3}}\left(\frac{a}{a_0}\right)^{3(1+w)/2}
\right]^{\gamma_{w3}/(\gamma_{w3}+1)}\\
&\sim \left(\frac{a}{a_0}\right)^{3(1+w)\gamma_{w3}/2(1+\gamma_{w3})}.
\end{split}
\eq
Even for the phantom, both the temperature and the entropy are positive.
For $w\ge -1$, during the DE domination, the radiation entropy
decreases as $a^{3(1+3w)/2}$,
the DE entropy decreases as $a^{3(1+w)\gamma/2(1+\gamma)}$
and the horizon entropy increases as $a^{3(1+w)}$, so the entropy bound
will be respected. As we discussed earlier, the entropies of the radiation and
DE with $w\ge -1$ increases slower than that of the apparent horizon,
so the entropy bound is always satisfied.  While for the phantom, $w<-1$,
both the horizon
entropy and the radiation entropy decrease and the phantom entropy
increases, the entropy bound will be violated at late times.

For the GCG, the solution to Eq. (\ref{theq5})
during RD is
\bq
\lb{rdtc5}
\frac{\gamma_{c1}\beta_{c1}}{\gamma_{c1}+1}T_c^{\gamma_{c1}+1}
=\frac{4\pi}{3}
\left(\frac{8\pi G}{3}\right)^{-3/2}(1+w_{c0})^{1/(1+\alpha)}
\rho_{r0}^{-3/2}\rho_{c0}\left(\frac{a}{a_0}\right)^3.
\eq
So the temperature scales as $T_c\sim a^{3/(1+\gamma_{c1})}$ and
the entropy scales as $S_c\sim a^{3\gamma_{c1}/(1+\gamma_{c1})}$. If
$\gamma_{c1}<-1$, then the entropy
increases faster than that of the radiation as the Universe expands.
If $\gamma_{c1}>0$, then both the temperature and the entropy increases
and the entropy increases slower than that of the radiation
as the Universe expands.
During the GCG domination,
the solution to Eq. (\ref{theq5}) is
\bq
\lb{chapsol5}
\frac{\gamma_{c2}\beta_{c2}}{\gamma_{c2}+1}T_c^{\gamma_{c2}+1}
=C_3+\frac{4\pi}{3}
\left(\frac{8\pi G}{3}\right)^{-3/2}\rho_c^{-1/2}\left(\frac{\rho_{c0}}{\rho_c}
\right)^{\alpha+1}\left[\frac{-2\alpha w_{c0}}{3+2\alpha}
+(1+w_{c0})\left(\frac{a_0}{a}
\right)^{3(1+\alpha)}\right],
\eq
where $C_3$ is an integration constant. At early times, $a\ll a_0$, the temperature
scales as $T_c\sim a^{3/2(1+\gamma_{c2})}$ and the entropy scales
as $S_c\sim a^{3\gamma_{c2}/2(1+\gamma_{c2})}$. At late times, $a\gg a_0$, both
the temperature and the entropy approach to a constant. To require that
the DE entropy is negligible compared to that of radiation
at least up to the present, we
find that $\gamma_{c2}>0$. So both the DE temperature and entropy
increase first and then decrease and the DE entropy increases
slower than that of the radiation.

\section{The generalized second law}
\label{gsl}

Now we proceed to discuss the GSL for the Universe with DE.
The geometric entropy associated with the apparent horizon is $S_A=\pi \tilde{r}^2_A/G$. Using Eq.
(\ref{rateq}), we get
\bq
\lb{sadot}
\dot{S}_A=3S_A H\frac{\rho_t+p_t}{\rho_t}.
\eq
As long as $\rho_t+p_t\ge 0$, $\dot{S}_A\ge 0$ ,
while for the phantom domination, $\dot{S}_A<0$.

\subsection{Constant co-moving entropy density}

The radiation entropy contributed by the
relativistic particles inside the apparent horizon is $S=4\pi s \tilde{r}^3_A/3$,
its time derivative reads
\bq
\lb{desdot}
\dot{S}=3S H\frac{\rho_t+3p_t}{2\rho_t}.
\eq
In the radiation or matter dominated eras, $S$ increases as the
Universe expands if $S>0$, so the GSL is satisfied. The
violation of the GSL is only possible when
DE dominates. During the DE domination,
the variation of the total entropy in the Universe is expressed as
\bq
\lb{sandsa}
\dot{S}+\dot{S}_A=3S H\frac{\rho_t+3p_t}{2\rho_t}+
3 S_A H\frac{\rho_t+p_t}{\rho_t}.
\eq
During RD and MD eras, the radiation entropy increases
slower than that of the apparent horizon,
so the entropy bound $S < S_A$ is always satisfied.
Considering the equation of state in the RD and MD epoches, Eq. (\ref{sandsa})
is always positive so that the GSL is protected during RD and MD eras.
When the DE starts to dominate,
$S$ decreases as the Universe expands while $S_A$ keeps on increasing for $w>-1$,
so in the far future $S\ll S_A$ for $w>-1$.  Since $(\rho_t+3p_t)/2\rho_t$ has the same order of magnitude as
$(\rho_t+p_t)/\rho_t$, the GSL can be preserved for $w>-1$. To be more explicit, we take
the DE model with constant $w$ as an example. In this
case, Eq. (\ref{sandsa}) becomes
\bq
\lb{sandsa1}
\dot{S}+\dot{S}_A=3H\left[\frac{1+3w}{2}S_0\left(\frac{a}{a_0}\right)^{3(1+3w)/2}
+(1+w)S_{A0}\left(\frac{a}{a_0}\right)^{3(1+w)}\right].
\eq
For $w>-1$, $|(1+3w)/2|S_0<(1+w)S_{A0}$ and the first term in the right hand side
decreases and the second term increases, so the GSL is always
satisfied.

For the phantom with $w<-1$, from the discussion in Sec. \ref{fixw},
we learnt that the temperature $T<0$ while the entropy $S>0$. The GSL
is violated since
$\dot{S}_A<0$ and $\dot{S}<0$ during the phantom domination.
This problem can be attributed to the negative temperature
deduced in the formalism where the volume and the
temperature are assumed to be independent. In \cite{phanthermo},
by identifying the temperature to the apparent
horizon temperature, $T>0$ and $S<0$, it was shown that
the GSL may be satisfied.

\subsection{Volume as a function of Temperature}

Now we study the GSL by considering the physical volume
and the temperature are dependent and taking the temperature of the Universe
to be equal to the apparent horizon temperature.
From Eq. (\ref{theq2}), we see that $\dot{S}\ge 0$ ($<0$) if $(\rho+p)/T\ge 0$ ($<0$) during
RD and MD,  and $\dot{S}\le 0$ ($>0$) if $(\rho+p)/T\ge 0$ ($<0$) during the DE domination.
During the phantom domination, $\dot{S}_A<0$, so we must require $T>0$ to protect the GSL.

If $T=T_A$, Eq. (\ref{theq3}) tells us that
\bq
\lb{eqseq1}
\dot{S}=\frac{9}{16 G^2}(\rho_t+3p_t)(\rho+p)\rho_t^{-3}H.
\eq
Combining Eqs. (\ref{sadot}) and (\ref{eqseq1}), we get
\bq
\lb{eqgsl1}
\dot{S}_t+\dot{S}_A=\frac{27}{16G^2}H\rho_t^{-3}(\rho_t+p_t)^2\ge 0
\eq
for the total entropy of the Universe and the apparent horizon entropy.
Therefore, the GSL is always satisfied even for the phantom.

For the case of $S=\beta T^\gamma$, we find that the entropies all increase
during RD and MD. Only when the DE dominates, the entropies of the
radiation and DE with $w> -1$ decrease. So we discuss the GSL during
the DE domination. All the entropies have the power law dependence on
the scale factor $a$, so $\dot{S}\sim SH$ with an order one proportionality
factor. Since $S_R\ll S_A$ and $S_w\ll S_A$, so the GSL is always respected.

\section{Conclusions and Discussion}
\label{conc}

In this paper we have studied the thermodynamical properties of the Universe with DE.
Adopting the usual assumption in deriving the constant co-moving entropy density that the physical
volume and the temperature are independent, we studied the thermodynamics of
the Universe in the radiation, matter and DE dominated
periods. During RD and MD epoches, the entropy of the Universe
contributed by the relativistic particles increase with the expansion,
while in the DE dominated era, the entropy remains
positive but decreases as the Universe expands. In all
these processes, the entropy within the Universe is bounded
by the geometric entropy associated with the apparent horizon. The
holographic principle is respected. The requirement of
the holography at early epoch gives reasonable
energy scale. We have found some strange behaviors of the
DE based on the assumption that
the physical volume and the temperature are independent.
For the DE with constant $w$, we have observed the strange dispersion
relation $E\sim P^{3w}$. For the phantom, we saw that either the
temperature or the entropy could be negative. For the DE with constant
equation of state $w>-1$, the temperature increases as the Universe expands.
We have also extended our discussion to the GCG and found that
its temperature decreases as the Universe expands,
which is different from the DE with constant $w$.
Thus the strange behavior on the temperature as observed is not the general
property of the DE, it looks model dependent.

We have also considered the realistic situation that the physical volume
and the temperature of the Universe are related. We have concentrated
on the volume within the apparent horizon.
The apparent horizon is important for the study of cosmology,
since on the apparent horizon there is the well known
correspondence between the first law of thermodynamics
and Einstein equation. On the other hand it has been found that the apparent horizon
is a good boundary for keeping thermodynamical laws \cite{wang06}.
Considering that the Universe is in thermal equilibrium, $T=T_A$,
we have studied the entropy of the DE in the radiation, matter and
DE dominated eras. We found the DE entropy increases during the RD and MD eras while
decreases during the DE dominated era. In all epoches, the DE
entropy is bounded by the holographic entropy on the apparent horizon.
Requiring that at the present epoch, the DE entropy is negligible
compared to the radiation entropy, we can limit
the DE equation of state to be very close to $-1$,
which is consistent with observations. We found that both temperature and
entropy can be positive for the DE including the phantom.

We have also investigated the GSL in the Universe with DE.
By assuming the physical volume independent of the temperature,
we have found that the GSL is protected for the DE with $w>-1$.
Considering the realistic case
that the physical volume and the temperature are related, and the
DE inside the Universe is in thermal equilibrium with the boundary by the
apparent horizon, the GSL is proved to be always satisfied.

The results tell us that in studying the thermodynamics of the Universe
with DE, it is more appropriate to consider
the case that the physical volume and the temperature are related.
This could give us more reasonable results on the thermodynamical
quantities in the Universe with DE.

\begin{acknowledgments}
YG is supported by Baylor University, NNSFC under grant
No. 10447008 and 10605042, CMEC under grant No. KJ060502 and
SRF for ROCS, State Education Ministry. BW 's work was partially supported by NNSF of China, Ministry of Education of China and
Shanghai Education Commission.
\end{acknowledgments}

\end{document}